\documentclass[]{aa}  

\usepackage{graphicx}
\usepackage{adjustbox}
\usepackage{txfonts}
\usepackage{natbib}
\usepackage{hyperref}

\usepackage{subcaption}
\usepackage{caption}     

\begin{document}

   \title{Cluster characteristics of Galactic open clusters}

   \author{K.~Neumannov{\'a}\inst{1}
      \and J.~Sup{\'i}kov{\'a}\inst{1,2}
      \and N. Faltov{\'a}\inst{1}
	 \and P.~Mondal\inst{1}	 
	 \and E.~Paunzen\inst{1}
   \and T.~Ramezani\inst{1}
	 \and G.~Sz{\'a}sz\inst{1}}

   \institute{Department of Theoretical Physics and Astrophysics, Faculty of Science, Masaryk University, Kotl\'{a}\v{r}sk\'{a} 2, 611 37 Brno, Czechia \\
	\email{474150@mail.muni.cz}
     \and CESNET, Gener\'{a}la Píky 430/26 160 00 Prague 6, Czechia}
   \date{}

\date{}
 
  \abstract
   {Star clusters are perfectly suited to study the formation, evolution and dynamics of stars and their host
   galaxy. For this, we need to estimate total masses, radii and shapes of star clusters. These parameters 
   strongly influence their fates in the host galaxies. However, there is no agreement about these
   characteristics in the literature. The published values widely disagree.}
   {We estimated total masses and cluster radii using a very conservative approach by calculating member
   masses and then summing them up. Furthermore, we investigated the three-dimensional characteristics 
   of open clusters and their limitations. }
   {We used a commonly used grid of evolutionary tracks for different photometric sources, including
   2MASS, $Gaia$, and Pan-STARRS. The interpolation method was tested using synthetic clusters
   built with ASteCA. For estimating the three-dimensional shapes, we used the distribution of the 
   well-known Galactic coordinates and inverse parallaxes. }
   {We estimated total masses and radii of about 7000 open clusters, seeing traces of tidal tails. Individual
   values can be used in the future for the next $Gaia$ data releases to get even more precise values. 
   The current DR3 allows only the study of three-dimensional shapes of up to 500\,pc. Beyond this 
   distance, we find a significant needle-like shape caused by the uncertainties in observed parallaxes.}
   {}


   \maketitle
%

\section{Introduction} \label{introduction} 

One of the most significant dynamical characteristics of stellar populations is the initial 
mass function (IMF), the distribution of stellar masses at birth, which is generally assumed 
to be unchanging \citep{2026enap....2..173K}. Understanding many fundamental stellar processes,
including the first stars and galaxies origins and their development, heavily depends on the IMF.
There is no direct observational method for determining the star mass function (MF), although 
it is crucial in many areas of astrophysics. Thus, theories of stellar evolution are necessary 
to transform this observable quantity into the MF \citep{Chabrier2003,Sheikhi2016}.

Star clusters are groups of gravitationally bound stars which formed from the same 
molecular cloud, meaning they share certain characteristics, e.g., age, metallicity, 
distance, and reddening, thus providing an ideal environment for the study of star 
formation and evolution \citep{2019ARA&A..57..227K}. 

The total mass is one of the fundamental parameters of a star cluster. It determines its lifetime
and fate. A lower total mass means weaker gravitational binding, and therefore 
the loss of members due to gravitational nudges and the possible dissolving
within a few hundred million years \citep{2023MNRAS.524..968R}. Low-mass clusters are also
more affected by tidal stripping, i.e. tidal forces of the host galaxy \citep{2015ApJ...806..242R}.
For more massive star clusters, we find significant mass segregation, i.e. heavier stars 
sink to the centre while low-mass ones are pushed outward \citep{2007ApJ...655L..45M}.
They also then eventually undergo a process where the core becomes very dense, 
often forming central black holes \citep{2004ApJ...604..632G}. In additional, the collision and binarity
rates seem to depend on the total mass \citep{2025A&A...699A.142C}.

Several methods are used in the literature to estimate a cluster's mass. 
The most straightforward way is to calibrate the mass of all cluster members and to sum up
their masses. One of the most common ways to determine the mass of a single star is to use 
the measured luminosity. Many studies have determined the mass-luminosity relation (MLR) for
main-sequence stars in different mass ranges \citep{Eker_2015}. Here, one has to be careful
about binaries with unseen companions, the completeness of the member census, and the
limiting magnitude \citep{Piskunov2007}. It is also possible to apply the virial theorem, 
which gives the total mass from an estimate of
the stellar velocity dispersion and average interstellar distances \citep{2024A&A...686A..35T}.
It does not need knowledge of all cluster members. It very much depends on the
accuracy of the measured velocity dispersion \citep{2026ApJ...996..112M}.
Another method uses the tidal interaction of a cluster with its host galaxy. For this purpose,
\citet{1962AJ.....67..471K} introduced a core, tidal, and limiting radius together with the density
of star clusters. To calculate the individual total masses, a model for the gravitational field of the
host galaxy is needed. So, all methods have their advantages and disadvantages.

With the latest release of $Gaia$ DR3, researchers can dive deeper into the study of nearby open 
clusters \citep{2024A&A...686A..42H}. Currently, the kinematical characteristics and internal 
structures of these clusters remain poorly understood, as only a handful have been thoroughly
examined. The uncertainties in observed parallaxes have led to the clusters exhibiting elongated 
shapes rather than the more spherical forms one might expect, particularly when compared to 
findings from \citet{2021arXiv210707230P,Tarricq2022}. 
This motivated us to investigate the three-dimensional characteristics on the basis
of the currently available data and their errors.

In this paper, we present a comprehensive structural study of 7167 open clusters.

\section{Methods and analysis} \label{methods_analysis} 

For open cluster members, we used the following two very different lists.

\textit{\citet{2023MNRAS.526.4107P}\footnote{\url{https://ucc.ar/}}}: They used FASTMP (acronym for Fast Membership Probabilities)
to estimate membership probabilities. It can run in supervised and unsupervised modes.
It is based on pyUPMASK \citep{2021A&A...650A.109P}. 

\textit{\citet{2023A&A...673A.114H}}: They applied the Hierarchical Density-Based Spatial Clustering 
of Applications with Noise (HDBSCAN) algorithm to recover star clusters. 
They validated the found aggregates using a statistical density test and a Bayesian convolutional 
neural network for colour-magnitude diagram classification. 

We then checked the final list using the approach described in 
\citet[][Clusterix 2.0]{2020MNRAS.492.5811B}, which is based on proper-motion data using a 
fully non-parametric method. In an area occupied by a cluster, the frequency function 
is made up of two contributions: cluster and field stars. The tool performs an 
empirical determination of the frequency functions from the vector point diagram without 
relying on any previous assumption about their profiles. The comparison of all three 
methods yielded an excellent agreement.

For our study, we used the approach of estimating the mass of a cluster 
by summing up the masses of the individual members.
For this estimation, we used evolutionary tracks of appropriate ages, prepared
by \cite{Bressan2012PARSEC}, for different sources including 2MASS \citep{2006AJ....131.1163S},
$Gaia$, and Pan-STARRS \citep{2004SPIE.5489...11K}. 
To have a good estimation of the cluster's ages,  
the published values from the following five
references were averaged: \citet{2022MNRAS.509..421N}, \citet{2023A&A...673A.114H},
\citet{2024A&A...689A..18A}, \citet{2024AJ....167...12C}, and \citet{2025AJ....170..288L}.
For the member stars' mass estimation, we interpolated the masses of the nearest points in 
the evolutionary track\footnote{\url{https://github.com/Johaney-s/StIFT}}. 
This process was repeated for different colour-colour diagrams from 
the mentioned sources, and the average mass value was taken. After determining the masses 
of all member stars, we summed them up to obtain the total cluster mass. Additionally, we 
examined the influence of the restrictions placed on membership probability on the total mass.

We tested our procedure using the Automated Stellar Cluster Analysis package
\citep[ASteCA, ][]{2015A&A...576A...6P} using a grid of initial total masses, 
ages, and binary fractions within the $Gaia$ and 2MASS photometric systems. The initial 
parameters over the entire range could be reproduced with a scatter of about 5\%, respectively. 
This lends confidence in our method and the selected photometric systems.

\begin{figure}[h!]
    \centering
    \begin{minipage}{0.5\textwidth}
        \centering
        \includegraphics[width=\linewidth]{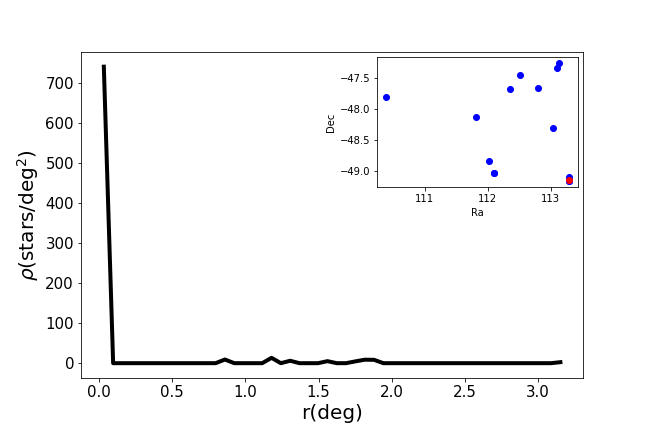}
        
    \end{minipage}
    \vspace{0.4cm} 
    \begin{minipage}{0.5\textwidth}
        \centering
        \includegraphics[width=\linewidth]{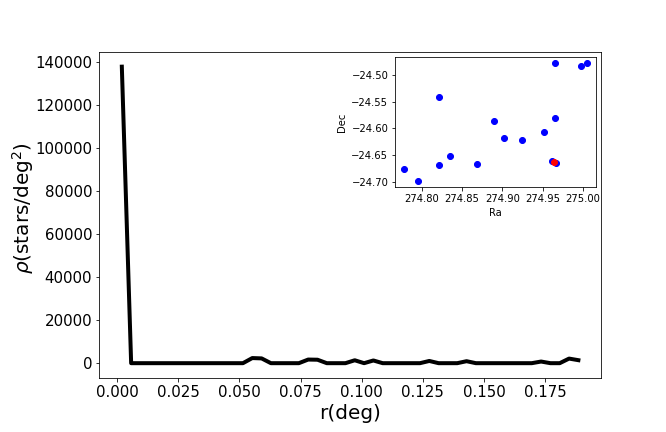}
        
    \end{minipage}
    \vspace{0.4cm} 
    \begin{minipage}{0.5\textwidth}
        \centering
        \includegraphics[width=\linewidth]{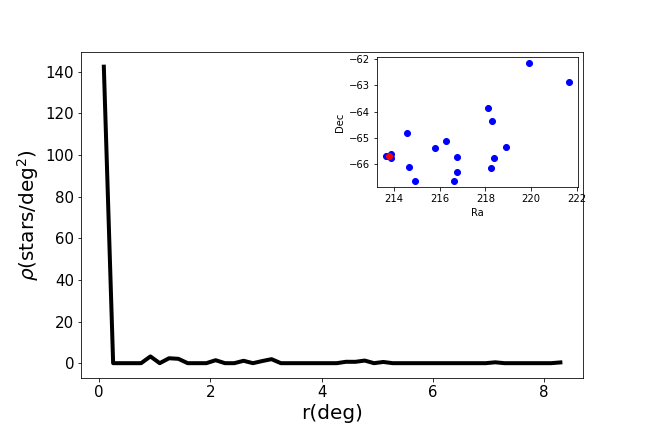}
        
    \end{minipage}
    \caption{Clusters CWNU 1265, HSC 109, and HSC 2631 are shown from top to bottom. The GOF values are all zeros. The stars are shown in blue, while the red dot indicates the centre of the cluster.}
    \label{fig:CWNU_1265_HSC_109_HSC_2631}
\end{figure}

\begin{figure}[h!]
    \centering
    \begin{minipage}{0.5\textwidth}
        \centering
        \includegraphics[width=\linewidth]{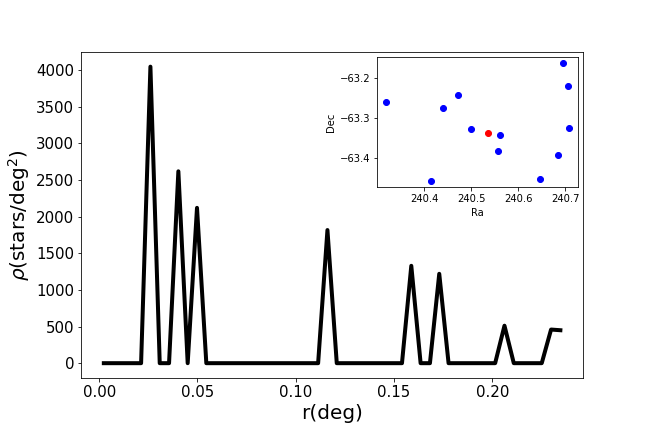}
        
    \end{minipage}
    \vspace{0.4cm} 
    \begin{minipage}{0.5\textwidth}
        \centering
        \includegraphics[width=\linewidth]{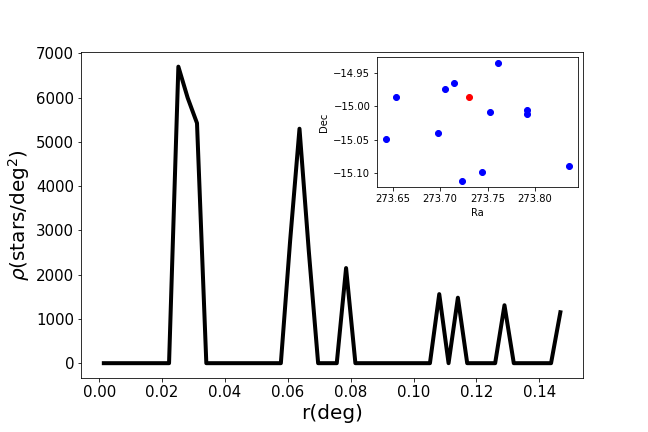}
        
    \end{minipage}
    \caption{Clusters HSC 2702 and HSC 184 are shown from top to bottom. The GOF values are as follows: 0.03 and 0.03. The stars are shown in blue, while the red dot indicates the centre of the cluster.}
    \label{fig:HSC_2702_HSC_184}
\end{figure}

\begin{figure}[h!]
    \centering
    \begin{minipage}{0.5\textwidth}
        \centering
        \includegraphics[width=\linewidth]{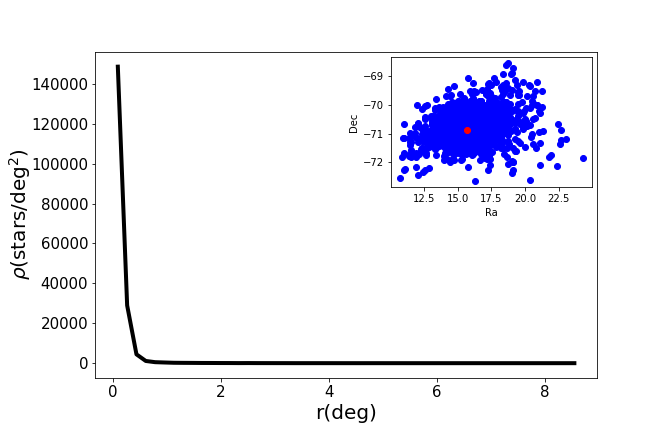}
        
    \end{minipage}
    \vspace{0.4cm} 
    \begin{minipage}{0.5\textwidth}
        \centering
        \includegraphics[width=\linewidth]{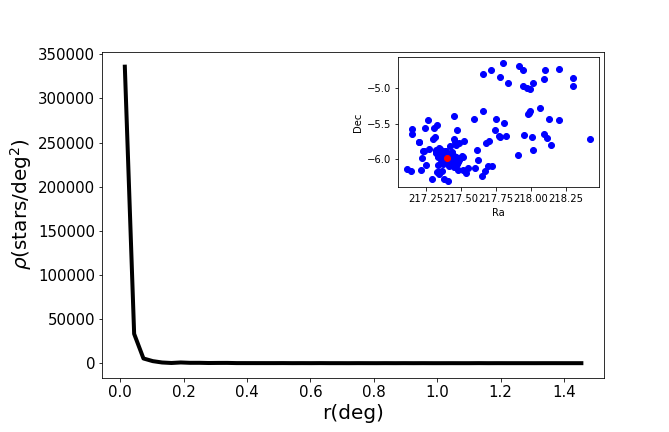}
        
    \end{minipage}
    \vspace{0.4cm} 
    \begin{minipage}{0.5\textwidth}
        \centering
        \includegraphics[width=\linewidth]{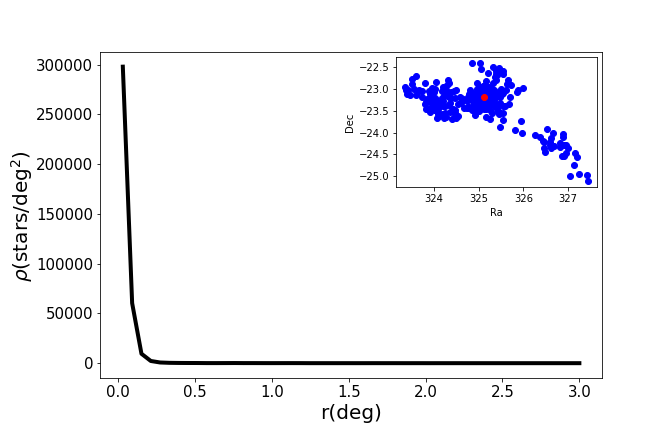}
        
    \end{minipage}
    \caption{Clusters NGC 362, NGC 5634 and NGC 7099 are shown from top to bottom. The GOF values are all higher than 0.999. The stars are shown in blue, while the red dot indicates the centre of the cluster.}
    \label{fig:NGC_362_NGC_5634_NGC_7099}
\end{figure}

\begin{figure}[h!]
    \centering
    \begin{minipage}{0.5\textwidth}
        \centering
        \includegraphics[width=\linewidth]{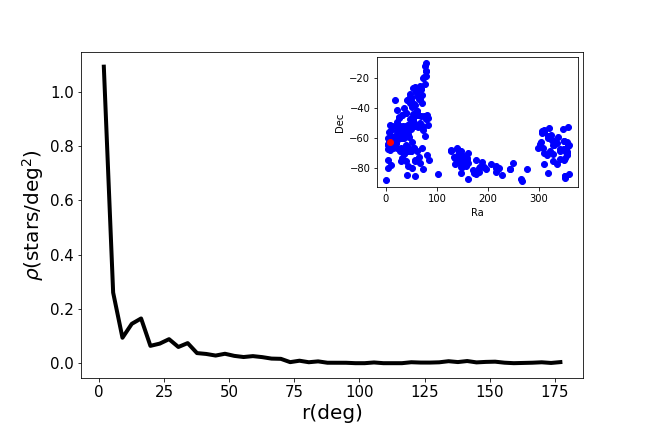}
        
    \end{minipage}
    \vspace{0.4cm} 
    \begin{minipage}{0.5\textwidth}
        \centering
        \includegraphics[width=\linewidth]{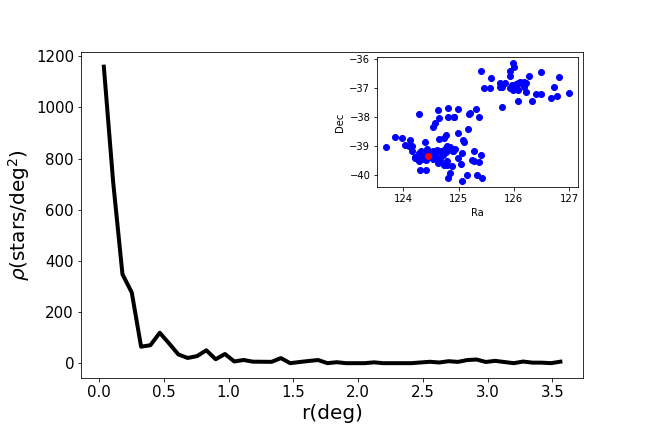}
        
    \end{minipage}
    \caption{Clusters Beta Tucanae Group and CWNU 338 are shown from top to bottom. The GOF values are as follows: 0.94 and 0.98. The stars are shown in blue, while the red dot indicates the centre of the cluster.}
    \label{fig:Beta_Tuc_CWNU_338}
\end{figure}

\begin{figure}[h!]
    \centering
    \begin{minipage}{0.5\textwidth}
        \centering
        \includegraphics[width=\linewidth]{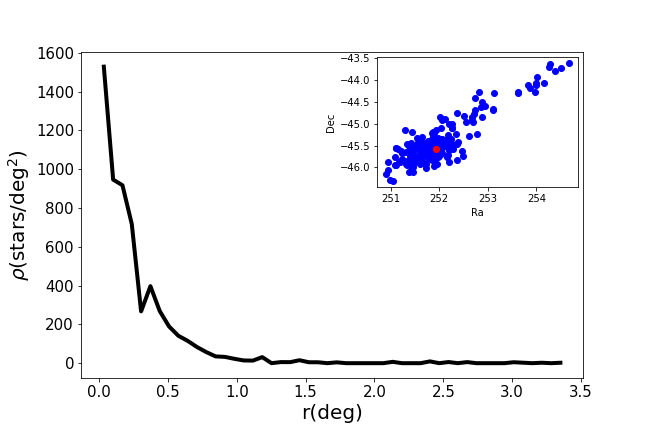}
        
    \end{minipage}
    \vspace{0.4cm} 
    \begin{minipage}{0.5\textwidth}
        \centering
        \includegraphics[width=\linewidth]{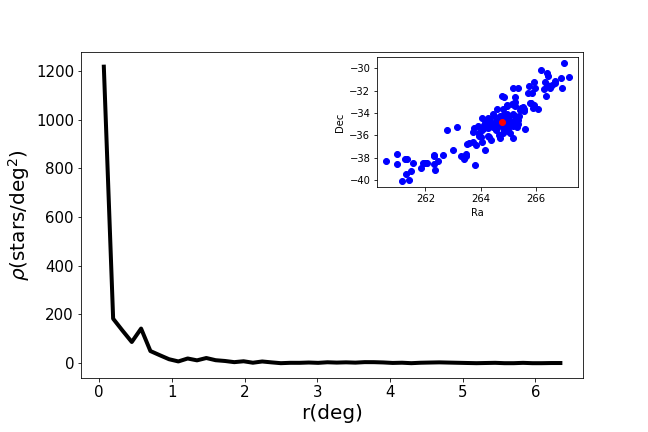}
        
    \end{minipage}
    \vspace{0.4cm} 
    \begin{minipage}{0.5\textwidth}
        \centering
        \includegraphics[width=\linewidth]{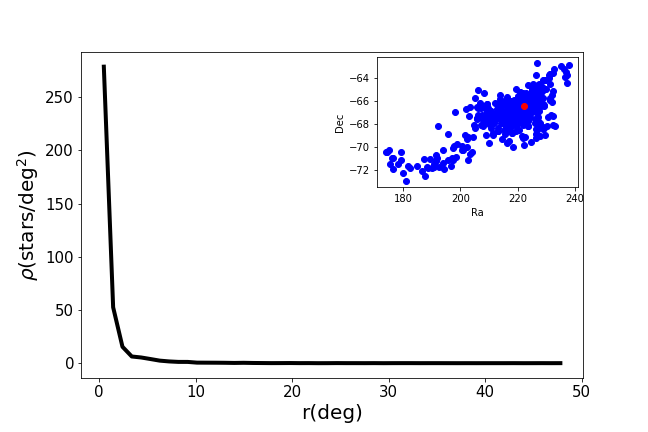}
        
    \end{minipage}
    \caption{Clusters ASCC 85, ASCC 90, and BH 164 are shown from top to bottom. The GOF values are as follows: 0.98, 0.97 and 0.99. The stars are shown in blue, while the red dot indicates the centre of the cluster.}
    \label{fig:ASCC_85_ASCC_90_BH_164}
\end{figure}

\subsection{Density profile}

The data were processed in Python to show the distribution of star density as a function of distance from the cluster centre. The data were first organised into two structures. The first associated each cluster with the coordinates Ra and Dec of each star in a given cluster and assigned them to the corresponding cluster, while the second included a list of cluster names for which the positions of the stars with their Ra and Dec coordinates were available. 

To correctly compute angular distances for clusters crossing the Ra = 0\degr\ boundary, coordinate wrapping was implemented. These conditions allow the calculation of distances between points located on opposite sides of the spherical coordinate boundary. 
To determine the density profiles, we used the method described in papers \citet{Zhong_2022} and \citet{BADAWY2024102196}, where we divided each cluster into concentric annuli (zones). 
The maximum stellar distance from the cluster centre was divided into N equally spaced radial bins.
The size of the step into which the circles are divided is given by the maximum distance of the star from the centre (the step between the radii of adjacent circles). We then calculated the number and densities of stars in each circle. Finally, we calculated the central radius of the circle ($\texttt{r\_centers}$)
as $ \rho(r) = N_i/A_i, A_i = \pi(r^2_{out} -r^2_{in}) $, which calculates the average distance of each circle from the centre (the midpoint between the inner and outer circles).

For the density profile, we assume the star density decreases exponentially with distance from the centre. Therefore, we present a model that is described by the exponential equation $ \rho(r) = a \cdot e^{-br} $
where $a$ is the central density and $b$ is the scale parameter, by which we determined the normalised goodness of fit (GOF). The GOF quantifies the degree of fit between the model and the data and allows us to evaluate how well the chosen model fits the true distribution of stars. The GOF ranges from 0 to 1, where 1 indicates that the model fully fits the data, and 0 indicates that the data are off. The GOF evaluates the entire profile and gives us a basic outline of whether a given star belongs to the system.

Thus, we have obtained a density profile for each cluster (the dependence of $\texttt{r\_centers}$ in units of deg on the density in units of  $ \mathrm{stars/deg^2} $ together with the best-fitting exponential model. At the same time, we stored the GOF value for each cluster and the parameters $a $ and $b $ in a text file. Finally, a spatial distribution map was generated for each cluster. Member stars are shown together with the cluster centre (red), with Ra and Dec plotted on the axes, allowing us to clearly identify the spatial distribution
of stars within the cluster.

\begin{figure}
    \centering
    \includegraphics[width =0.89 \columnwidth]{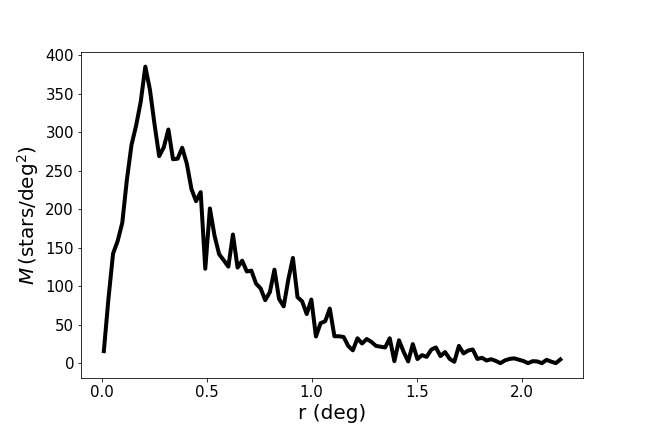}
    \caption{Mass profile for open cluster Theia 3623. We can notice the exponential function of the graph - the mass decreases with distance from the cluster's centre - Mass segregation.}
    \label{fig:Theia}
\end{figure}

\begin{figure}
    \centering
    \includegraphics[width =0.89 \columnwidth]{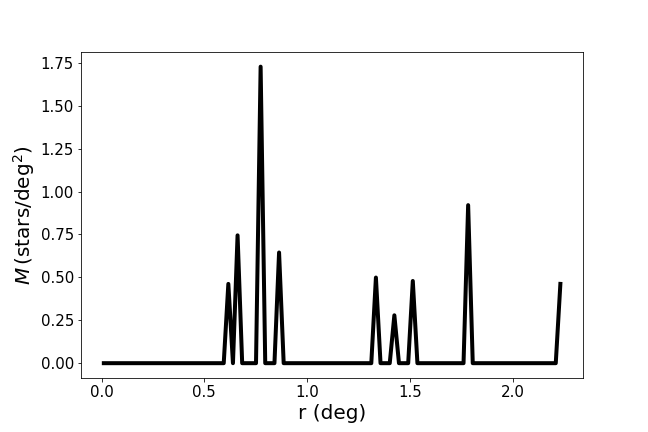}
    \caption{Mass profile for open cluster HSC 1941. The low number of members and the lack of evident mass segregation caused disorder in the mass profile.}
    \label{fig:HSC_1941}
\end{figure}

\begin{figure}[ht]
    \centering
    \subfloat[Alessi 43 - mass profile]{\includegraphics[width=0.45\textwidth]{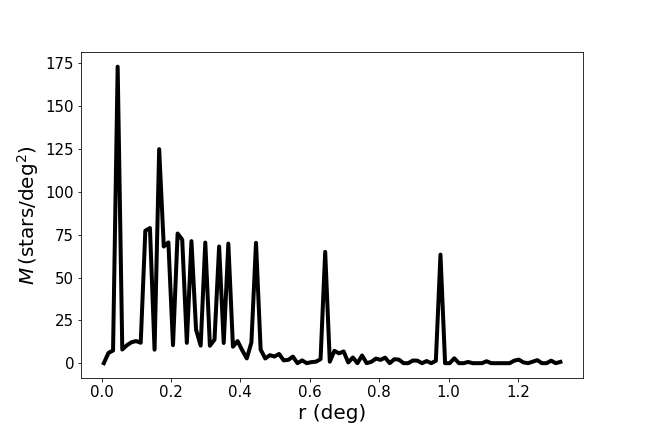}%
    \label{fig:Alessi_43}}
    \hfill
    \subfloat[Alessi 43 - density profile]{\includegraphics[width=0.45\textwidth]{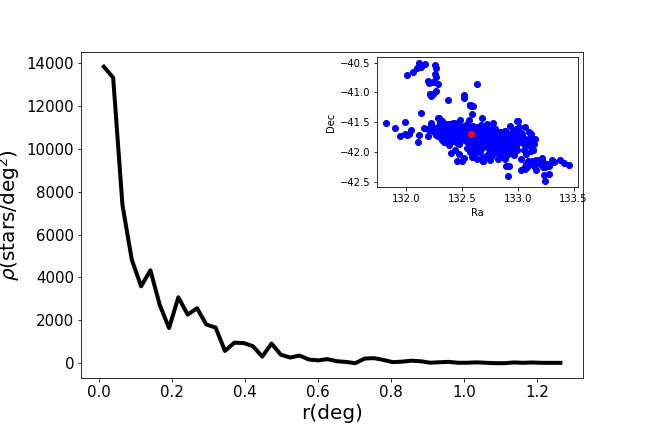}%
    \label{fig:Alessi_43_density}}
    \caption{Mass and density profile for open cluster Alessi 43. While the density profile shows an exponential decrease in density with distance, the mass profile does not. The blue points are members of the cluster, and the red dot is the cluster's centre.}
    \label{fig:Alessi_43_Pair}
\end{figure}

\begin{figure}[ht]
    \centering
    \includegraphics[width=0.89\columnwidth]{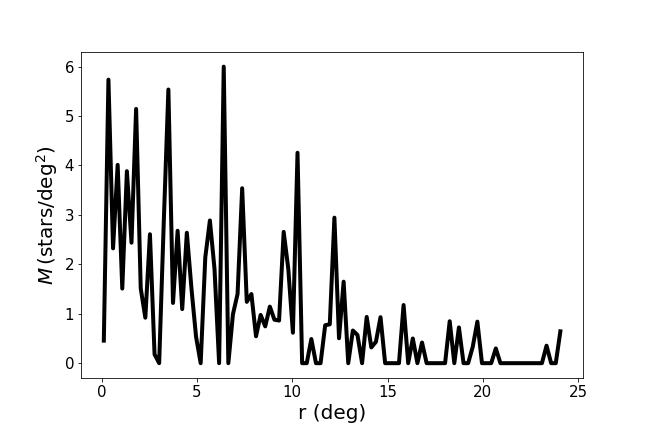}
    \caption{Mass profile for open cluster Alessi 13. The exponential decrease in mass with distance is unclear - not evident mass segregation for Alessi 13.}
    \label{fig:Alessi_13_Solo}
\end{figure}

\begin{figure}[ht]
    \centering
    \subfloat[NGC 6791 - mass profile]{\includegraphics[width=0.45\textwidth]{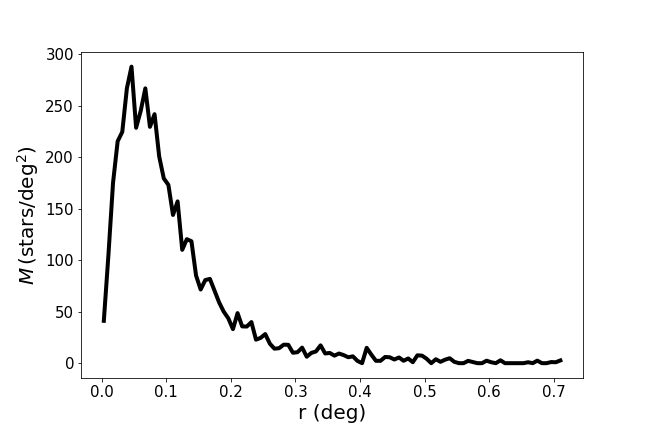}%
    \label{fig:NGC_6791}}
    \hfill
    \subfloat[NGC 6791 - density profile]{\includegraphics[width=0.45\textwidth]{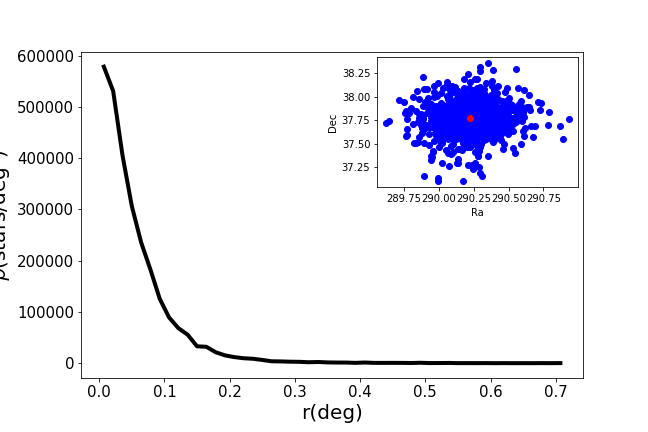}%
    \label{fig:NGC_6791_density}}
    \caption{Both mass and density profiles show an exponential profile with increasing distance. The cluster is well-organised - stars are distributed spherically around the cluster centre. }
    \label{fig:NGC_6791_Pair}
\end{figure}

\begin{figure}[ht]
    \centering
    \subfloat[CWNU 1044 - mass profile]{\includegraphics[width=0.45\textwidth]{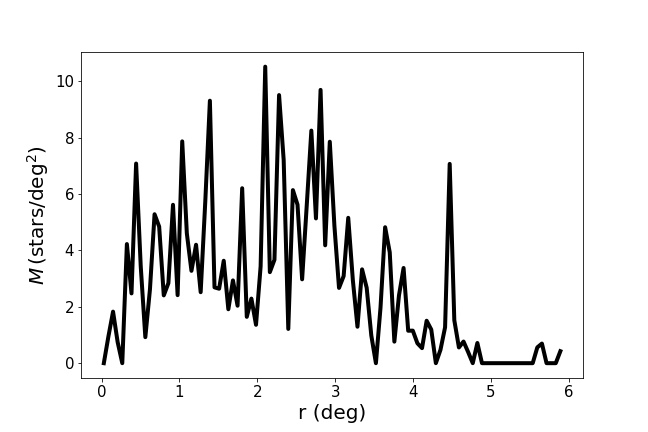}%
    \label{fig:CWNU_1044}}
    \hfill
    \subfloat[CWNU 1044 - density profile]{\includegraphics[width=0.45\textwidth]{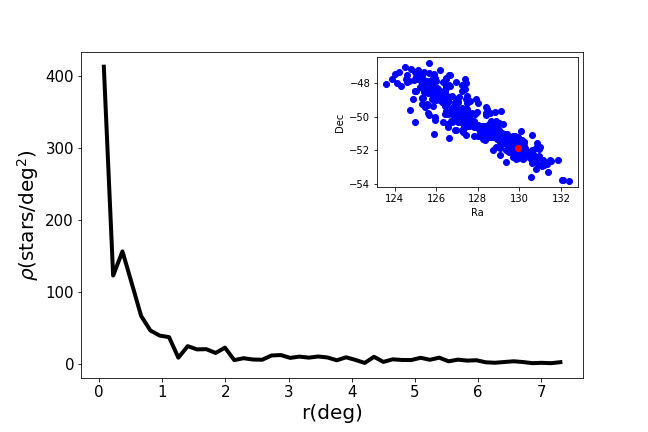}%
    \label{fig:CWNU_1044_density}}
    \caption{Density profile shows an exponential profile, while the mass profile does not. In the mass profile, there are two obvious mass peaks for 1 deg and 2.5 deg.}
    \label{fig:CWNU_1044_Pair}
\end{figure}

\begin{figure}[ht]
    \centering
    \subfloat[HSC 749 - mass profile]{\includegraphics[width=0.45\textwidth]{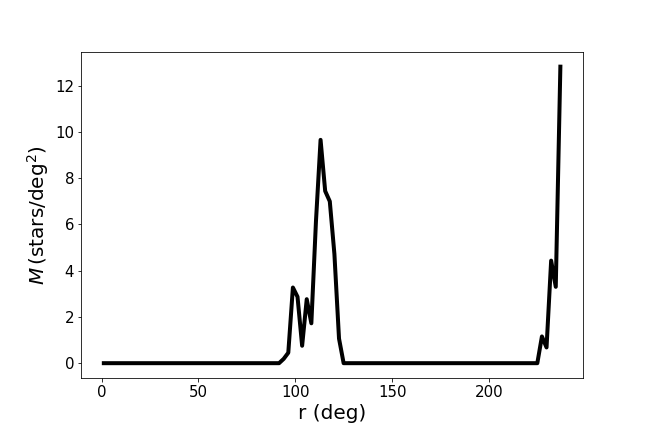}%
    \label{fig:HSC_749}}
    \hfill
    \subfloat[HSC 749 - density profile]{\includegraphics[width=0.45\textwidth]{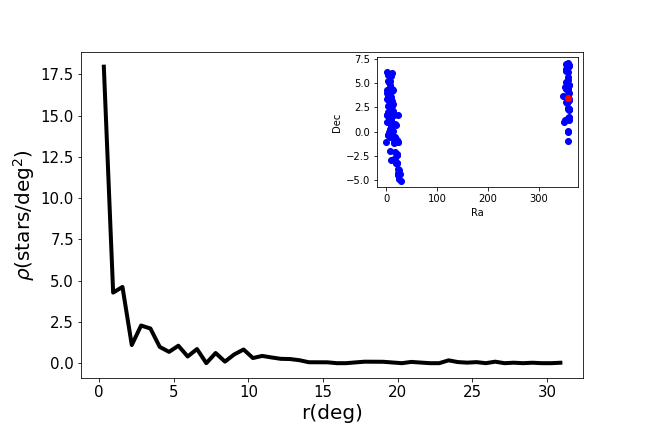}%
    \label{fig:HSC_749_density}}
    \caption{Two separated groups of stars caused two peaks in the mass profile.}
    \label{fig:HSC_749_Pair}
\end{figure}

\begin{figure}[ht]
    \centering
    \subfloat[HSC 939 - mass profile]{\includegraphics[width=0.45\textwidth]{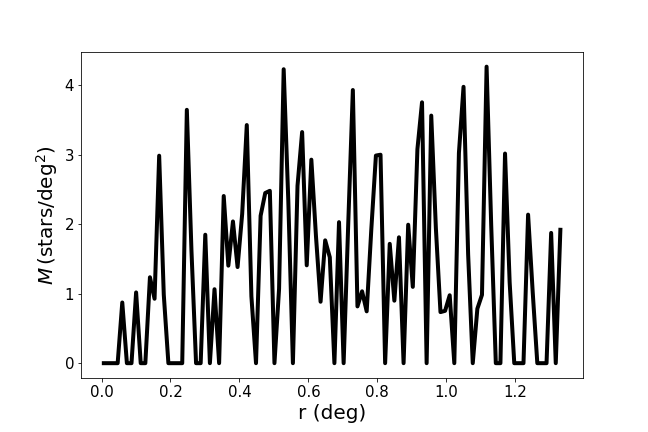}%
    \label{fig:HSC_939}}
    \hfill
    \subfloat[HSC 939 - density profile]{\includegraphics[width=0.45\textwidth]{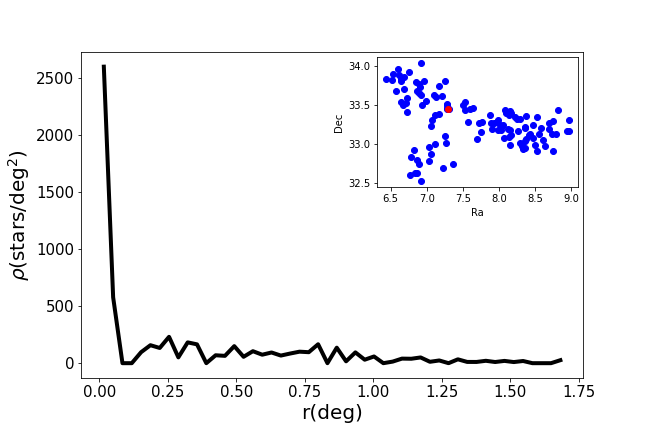}%
    \label{fig:HSC_939_density}}
    \caption{Stars in HSC 939 are located randomly into three unclear groups around the centre of the cluster. No mass segregation is observed.}
    \label{fig:HSC_939_Pair}
\end{figure}

\begin{figure}[ht]
    \centering
    \subfloat[NGC 6231 - mass profile]{\includegraphics[width=0.45\textwidth]{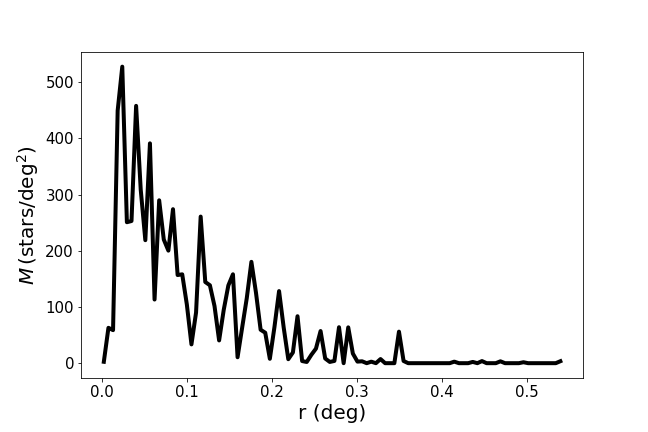}%
    \label{fig:NGC_6231}}
    \hfill
    \subfloat[NGC 6231 - density profile]{\includegraphics[width=0.45\textwidth]{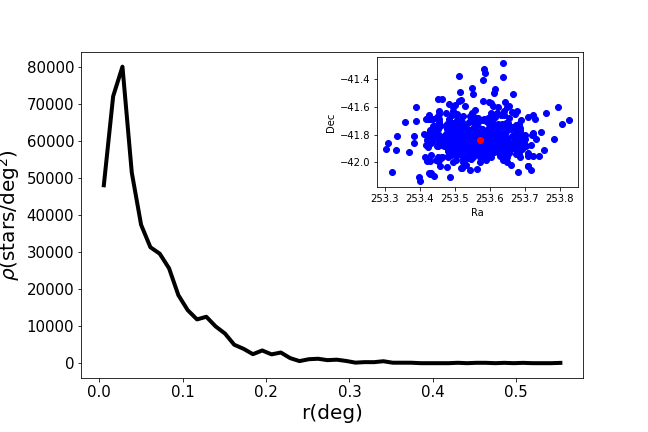}%
    \label{fig:NGC_6231_density}}
    \caption{Both graphs show exponential profiles thanks to the number of stars and their location around the centre of the cluster.}
    \label{fig:NGC_6231_Pair}
\end{figure}

\subsection{Mass profile} \label{analysis_masses}

To derive the stellar mass distribution as a function of distance from the cluster centre, the data were analysed using Python. To determine the mass profile, we 
followed the methods described in \citep{Zhong_2022} and \citep{BADAWY2024102196}.

First, for each star, its distance from the centre of the corresponding cluster was calculated. 
Then, the maximum distance of the star from the centre was found, based on which the cluster 
was divided into circles (zones). The maximum radial distance was divided into a predetermined number of equally spaced radial bins/annuli/circles.

Each star was subsequently assigned to the appropriate circles according to its distance from the 
centre. The masses of the stars that fell within each circle were summed. At the same time, the 
area between the circles was calculated, allowing the mass density in each interval to be determined 
as the ratio of the total mass in the interval to the area of the corresponding annulus/circles.

Finally, for each cluster, the total mass in each interval was plotted as a function of the distance 
from the central radius of each radial bin/annulus/circle. This plot shows how the mass of the stars is 
distributed within the cluster. The total mass of the cluster was calculated as the sum of the 
masses of all stars in each ring, with stars within the cluster varying in mass from about 
$10^4$ M$_\odot$ for the most massive stars in young clusters \citep{eaa_starclus} to less 
than 0.08 M$_\odot$. Within some clusters, mass segregation may exist, with the more massive 
stars concentrated towards the centre where the potential well is deeper.

\section{Results} \label{results}

In this section, we present the results for the various cluster parameters. 

\subsection{Density profile}

For our 7167 clusters, we plotted the density profile in a single graph and the coordinates (Ra and Dec) of each star (blue) and the cluster centre (red) in the upper-right corner of the same graph.

The largest GOF (over 0.999) was for the cluster NGC 362, which has a large number of stars, and most of them are close together near the centre, with only a few scattered. In total, 3782 clusters, i.e., about 53\% of the total, have a GOF greater than 0.9, making it almost certain that the exponential function perfectly describes the density; thus, the stars fall within the cluster.

In contrast, 1659 clusters (23 \%) have a GOF less than 0.5, of which 147 have a GOF equal to zero. These zero values are due to the algorithm failing to find optimal parameters to fit the model. There are several reasons: the function does not fit the nature of the data, and the values are too scattered.

Low GOF values are typically associated with clusters containing only a small number of stars and highly irregular density profiles. Although these stars may be part of the cluster, their low number distorts the GOF value. Thus, averaging into a single value may better capture the overall trend than the specified function. Conversely, for high GOF values, several strongly scattered stars may be present along with stars that are more centrally concentrated. This contrast often leads to a better GOF value compared to low GOF values, where there may be a small number of stars only slightly scattered.

Clusters with a zero GOF value, such as CWNU 1265, HSC 109, and HSC 2631, share a common feature - a low number of stars that could be considered as cluster members; see Fig. \ref{fig:CWNU_1265_HSC_109_HSC_2631}. Their density profile does not show the expected exponential decrease, but rather a steep, abrupt decline. Clusters with a very low but non-zero GOF exhibit inconsistent density profiles, characterised by multiple local maxima. These peaks contrast with the otherwise mostly horizontal profile, see Fig. \ref{fig:HSC_2702_HSC_184}. This phenomenon suggests that the distribution of stars in these clusters is highly irregular and probably affected by dynamical processes or small sampling statistics.
In contrast, NGC 362 (highest value), NGC 5634, and NGC 7099 have the highest GOF (over 0.999), see Fig. \ref{fig:NGC_362_NGC_5634_NGC_7099}.

The Beta Tucanae Group shows a very unusual distribution of stars compared to classical globular or elongated cluster shapes. Despite having a high GOF value (0.93), the stars in this group are distributed in three separate regions. In addition, one of these regions is characterised by a distinctly vertically elongated shape, see Fig. \ref{fig:Beta_Tuc_CWNU_338}. These separated regions may indicate that the stars were formed in different regions of the molecular cloud or that they were affected by gravitational interactions with other stars or nearby matter.

Interactions with the surrounding environment, such as passing through a denser region of the galaxy or encountering other clusters, can also split stars into separate groups. The elongated shape of one region may be the result of dynamical interactions or gravitational interactions with nearby objects, where, as we can see, there is a higher concentration of stars in that region.

The cluster CWNU 338 shows a similar elongated shape. The elongated shape, most likely caused by tidal forces, is also present in clusters ASCC 85, ASCC 90, and BH 164, where all of these clusters also have high GOFs (more than 0.97), see Fig. \ref{fig:ASCC_85_ASCC_90_BH_164}.

\subsection{Mass profile}

Our input data contained stars falling into 7168 open clusters, while stellar mass data were available only for stars in 6926 clusters. Stellar mass information was provided as the stellar mass percentile for a single star, in units of solar mass (M$_\odot$). For each cluster, the following percentiles were available: 2.5, 16, 50, 84 and 97.5.

Each cluster was analysed for all given percentiles, and the total mass of the cluster corresponding to each percentile calculated. These total masses represent the sum of the masses of the stars falling in that percentile (e.g., for percentile 97.5, it is the sum of the masses of the lowest-mass 97.5 \% of the stars in that cluster – only 2.5 \% of stars are more massive).

For the percentiles 50, 84, and 97.5, we identified the same cluster as both the most and the least massive. The most massive cluster was Theia 3623 with a mass of 9019 M$_\odot$, while the least massive cluster was HSC 1941 with a mass of 6.2 M$_\odot$ (value corresponding to percentile 97.5). 
For the 2.5th and 16th percentiles, we also identified the same cluster as both the most and least massive. The most massive was HSC 655 with a mass of 4714 M$_\odot$, while the least massive was HSC 940 with a mass of 1.9 M$_\odot$ (value corresponding to the 2.5th percentile).

The total mass of the cluster decreases with lower percentiles because the lower percentiles include lower-mass stars, while the higher percentiles are dominated by more massive stars, which increase the average mass. Although the number of stars across all percentiles remains constant, the lower percentiles correspond to stars with lower masses, while the higher percentiles include stars with higher masses, leading to higher total masses in those percentiles.

The cluster Theia 3623 contains 3 086 members. For percentile 97.5, the minimum mass of a member star is 1.7 M$_\odot$, and the maximum mass reaches 3.9 M$_\odot$. In contrast, the HSC 1941 cluster has only 10 members, with a minimum mass of 0.27 M$_\odot$ and a maximum mass within the same percentile of 1.7 M$_\odot$. The curves in Figs. \ref{fig:Theia} and \ref{fig:HSC_1941} showing the dependence of mass on distance differ significantly between the two clusters. For HSC 1941, the expected exponential decrease in mass with distance is not observed due to the low number of members, whereas this characteristic is well evident in Theia 3623.

The most massive member, with a mass of 63.1 M$_\odot$, was identified in the cluster Alessi 43, which has 519 members and a total mass of 1 526 M$_\odot$ (at the 97.5 percentile). Such a high stellar mass is consistent with a very massive early-type star. The Fig. \ref{fig:Alessi_43} shows the mass versus distance for this cluster, which shows an irregular distribution of stars by mass. This may be because it is a young object where mass segregation has not occurred, or, conversely, because it is very far apart and not interacting in any way. The second possibility is less likely, since the density profile for the cluster and the distribution of stars in the cluster in Fig. \ref{fig:Alessi_43_density} show us that the density profile decreases exponentially with distance and the stars are not dispersed in any way. So the first option seems more likely, with the WEBDA catalogue \citep{coascc} putting the age of the cluster at about 30 Myr. 

At the other end of the spectrum is the Alessi 13 cluster, which has 167 members. The lowest-mass star in this cluster has a mass of 0.09 M$_\odot$, and we expect that it is near the hydrogen-burning limit. The total mass of the cluster (for percentile 97.5) is 101.9 M$_\odot$, with the most massive member having a mass of 4.77 M$_\odot$. Considering the number of members and the total mass of the cluster, we can conclude that the cluster contains mostly very low-mass stars, and even the most massive stars have rather low masses. As can be seen in Fig. \ref{fig:Alessi_13_Solo}, the mass profile is also random in this case, suggesting that the stars are not systematically distributed according to mass. This phenomenon can be attributed to the fact that most of the stars have very low masses, only a small number of stars are more massive, and the sample of stars analysed is not very large.

The cluster with the largest number of members for which mass data were available is NGC 6791, containing 4 193 members with a total mass of 4 870 M$_\odot$ (for a percentile of 97.5). The star with the lowest mass has a mass of 0.81 M$_\odot$ and the most massive one has a mass of 1.86 M$_\odot$. The cluster contains mostly low and intermediate-mass stars. The mass profile, see Fig. \ref{fig:NGC_6791}  shows an exponential decrease in mass with distance, indicating strong mass segregation, supported by the age of the cluster (about 4 GYR according to the WEBDA catalogue). The density profile and the distribution of stars indicate a globular structure with a high concentration of stars in the centre of the cluster, see Figure \ref{fig:NGC_6791_density}.

Interesting mass and density profiles can be observed for the clusters CWNU 1044, HSC 749 and HSC 939, whose properties are as follows:

CWNU 1044: 377 members, total mass 272 M$_\odot$ (for a percentile of 97.5). The Fig. \ref{fig:CWNU_1044_Pair} shows an irregular increase and decrease in mass with distance, probably due to the elongated shape of the cluster.

HSC 749: 146 members, total mass 70 M$_\odot$. Both the density and mass profiles reveal two separate regions containing stars, as confirmed by the distribution of stars by coordinates, see Figure \ref{fig:HSC_749_Pair}.

HSC 939: 122 members, total mass 129 M$_\odot$. Here, no clear radial trend is observed; the stars have similar masses. The density profile shows three separate clusters of stars near the cluster centre, see Figure \ref{fig:HSC_939_Pair}.

The cluster NGC 6231 shows the largest difference between the number of members and the total mass, containing 646 stars with a total mass of 7041 M$_\odot$. The average mass of a single star is thus approximately 11 M$_\odot$. This cluster, therefore, contains mostly very massive stars, and given its low age (2-7 Myr according to \citep{Kuhn_2017}), many of the massive stars are expected to remain on or close to the main sequence. Although its mass profile shows an overall exponential decline, the sharp fluctuations suggest the presence of less massive stars alongside the significantly more massive ones. The density profile of this cluster shows an exponential decrease, indicating a concentration of stars, with an approximately symmetric distribution (see Figure \ref{fig:NGC_6231_Pair}).

\begin{figure*}[t]
   \centering
	 \begin{tabular}{cc}
   \subfloat[][The 3D structure of NGC 1039 at $d$\,=\,509(4)\,pc. \label{NGC_1039}]{\includegraphics[scale=0.5]{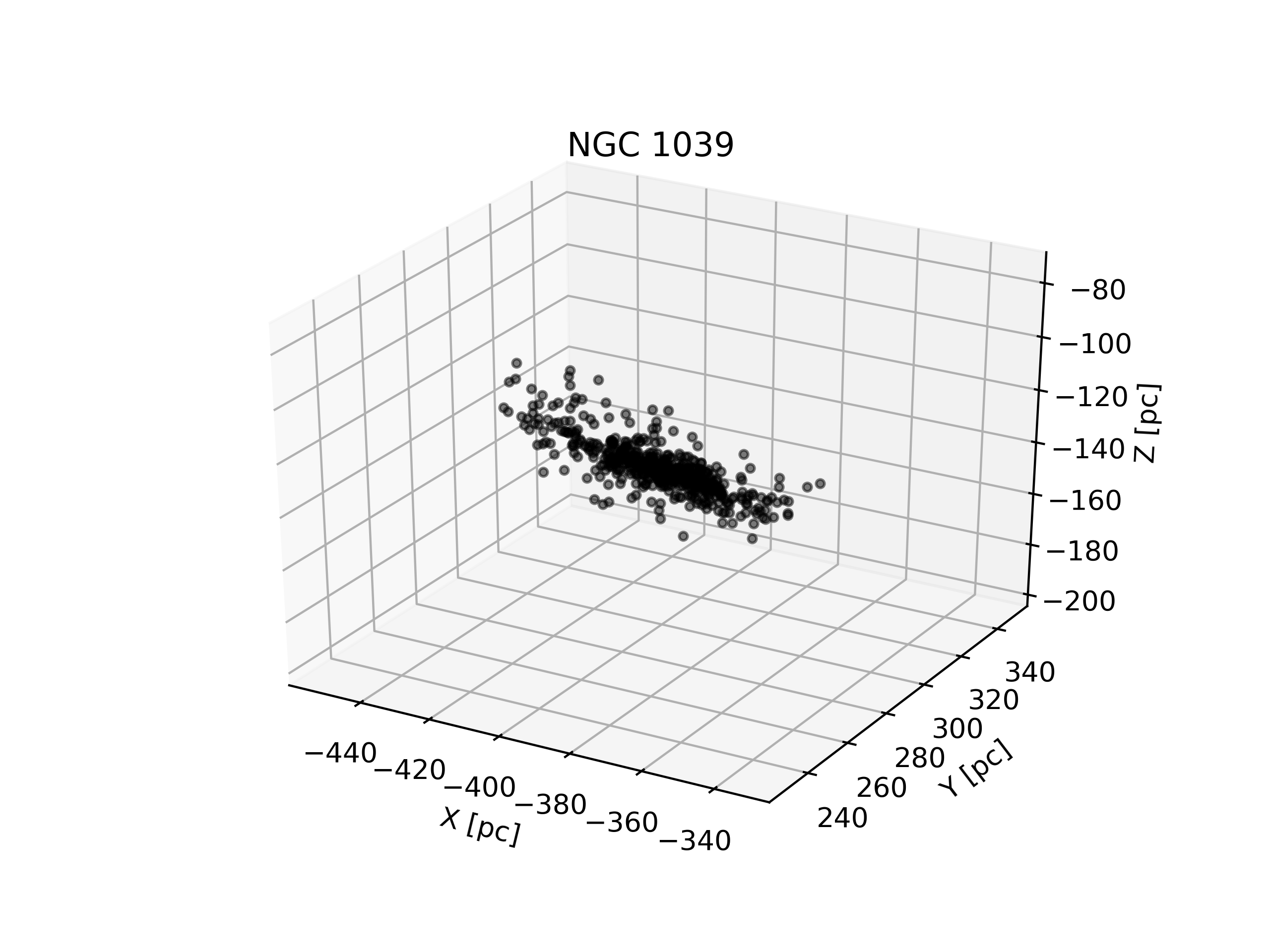}} \quad &
	 \subfloat[][The 3D structure of NGC 1528 at $d$\,=\,1038(6)\,pc. \label{NGC_1528}]{\includegraphics[scale=0.5]{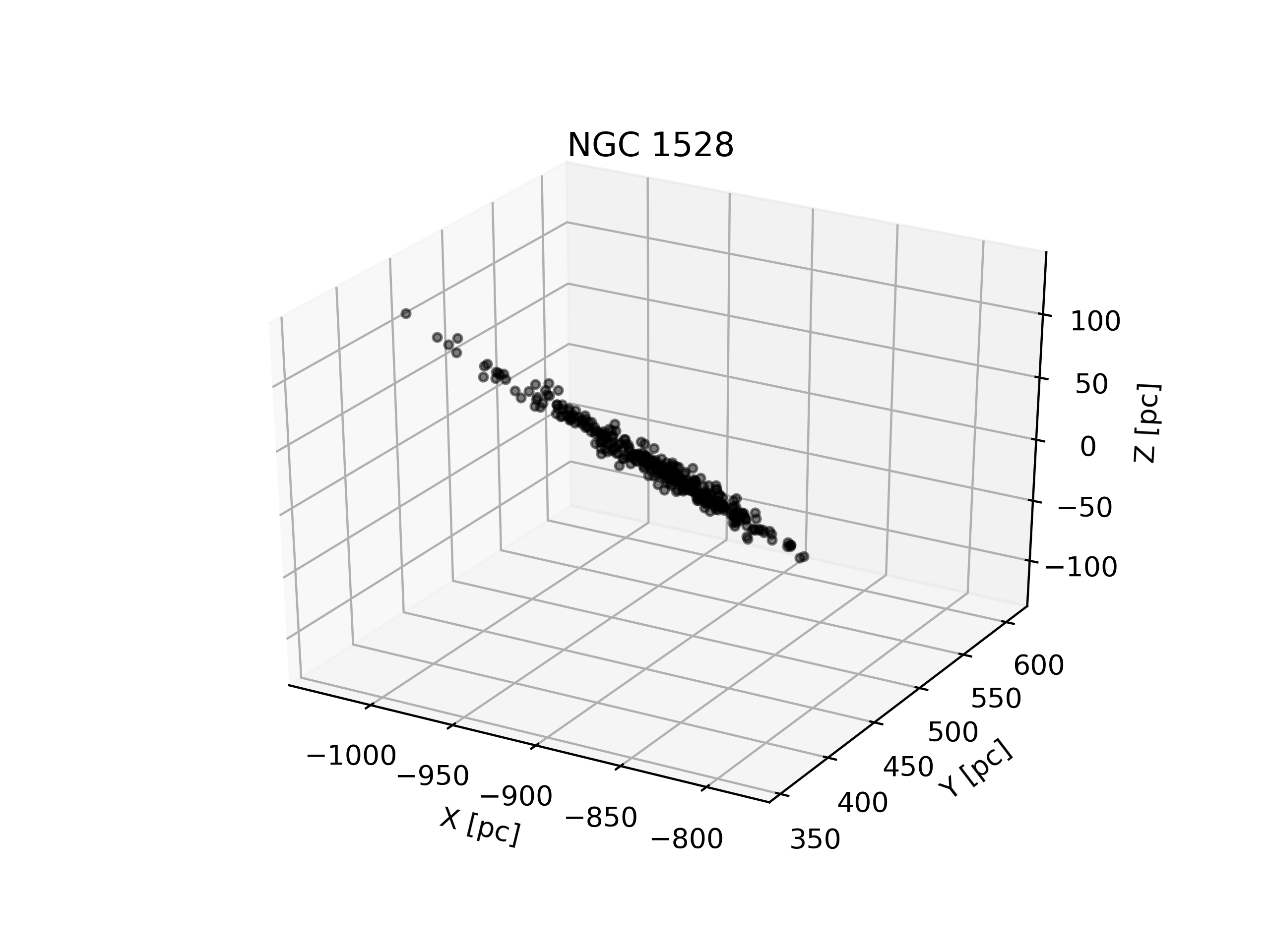}} \\
   \subfloat[][The 3D structure of NGC 2632 at $d$\,=\,192($1$)\,pc. \label{NGC_2632}]{\includegraphics[scale=0.5]{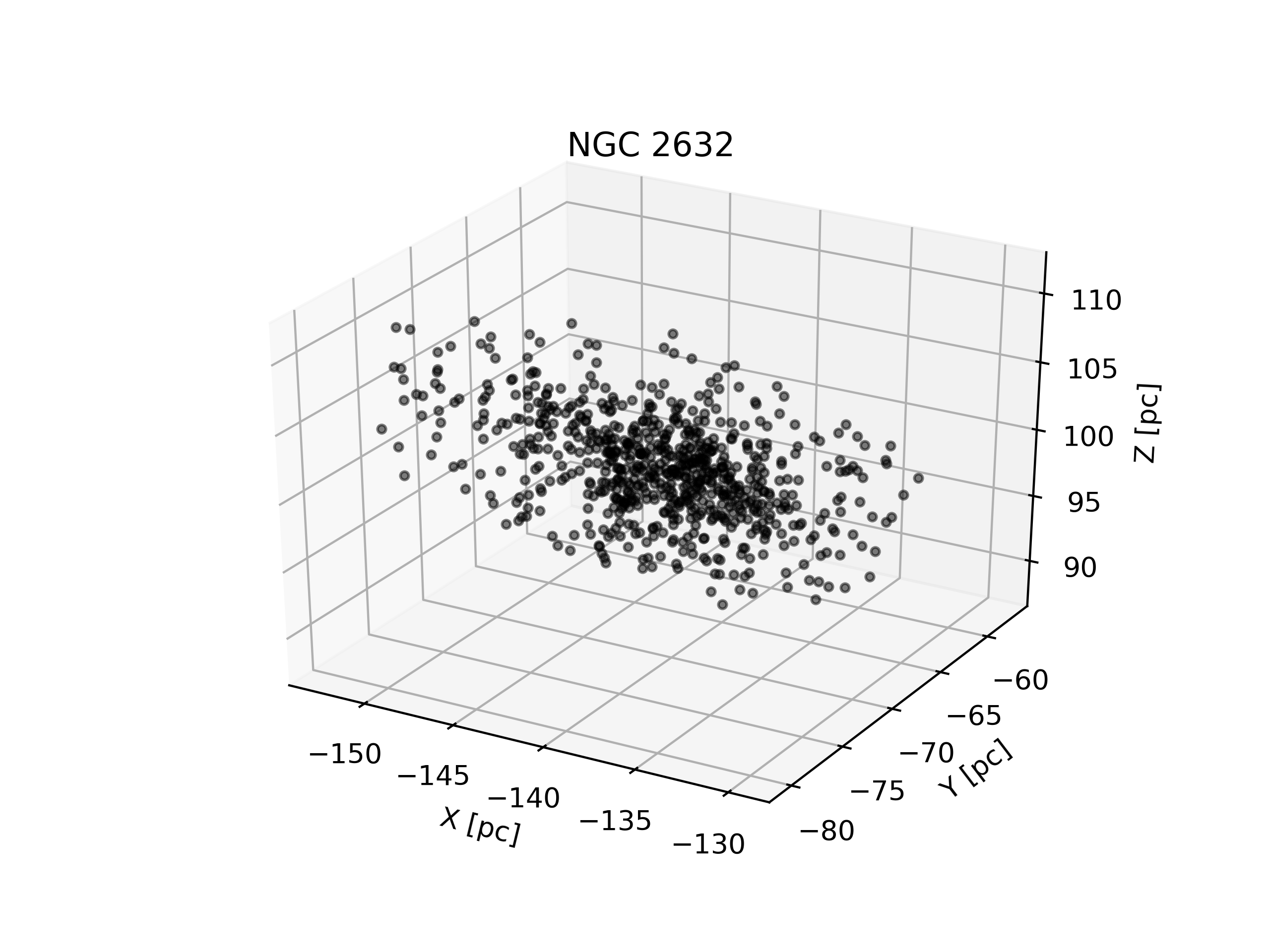}} \quad & 
   \subfloat[][The 3D structure of NGC 5823 at $d$\,=\,1756(14)\,pc. \label{NGC_5823}]{\includegraphics[scale=0.5]{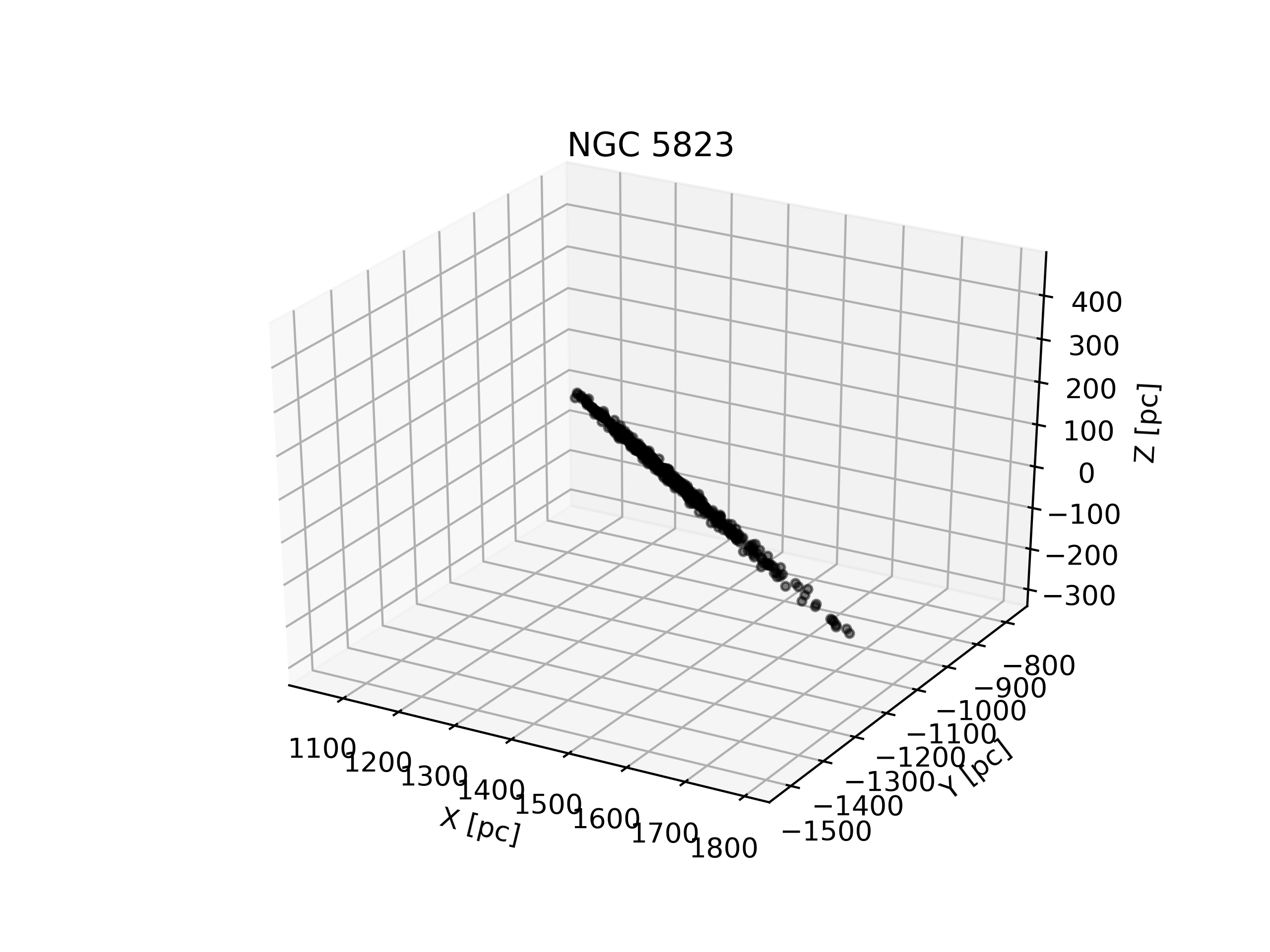}} \\
	 \end{tabular}
   \caption{The expected needle-like structure in the line-of-sight is clearly visible for more distant clusters.}
   \label{3D_four_clusters}
\end{figure*}

\begin{figure}
\centering
\includegraphics[width=0.45\textwidth]{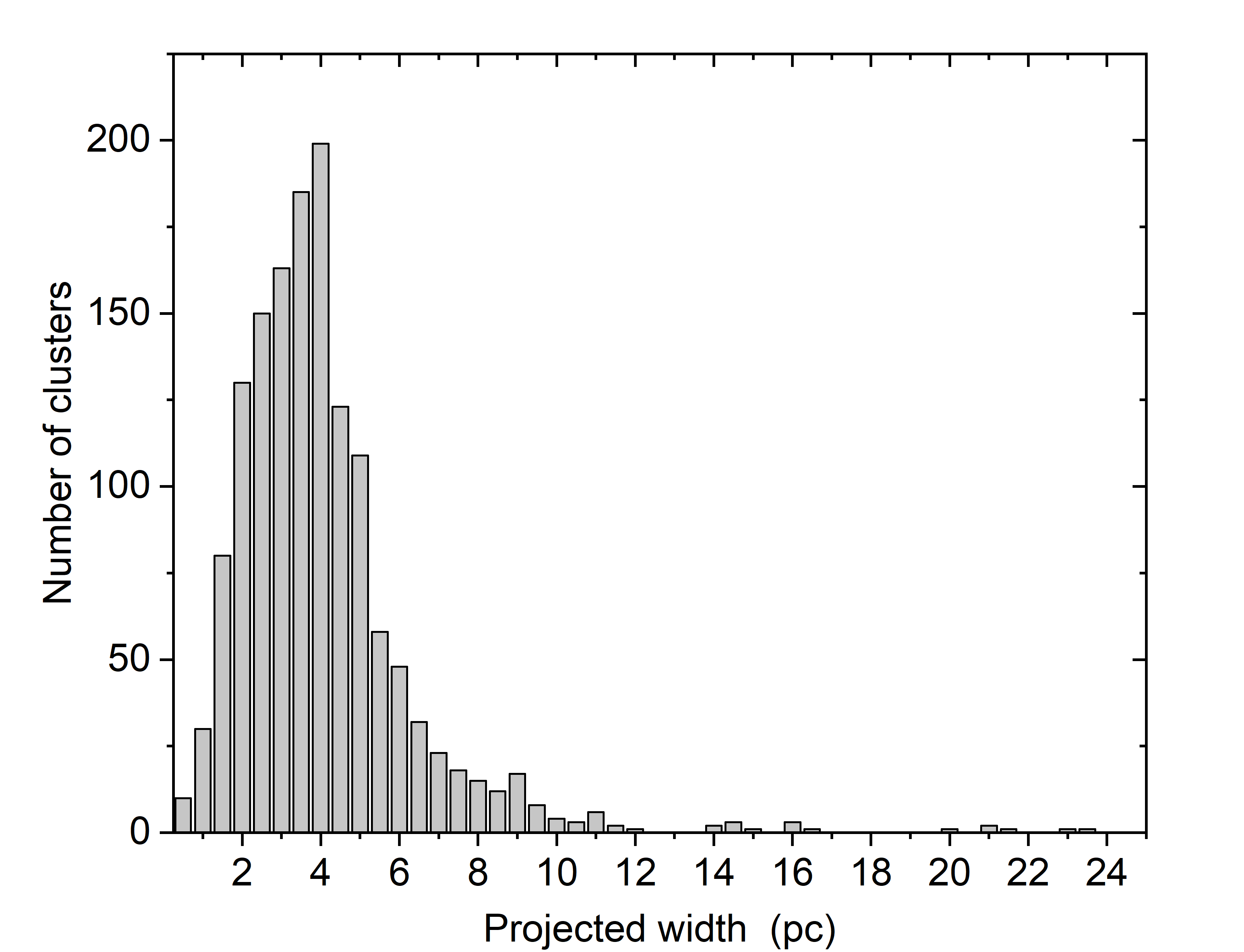}
\caption{The histogram of the projected widths. We can interpret 
these widths as a measure of the cluster size or diameter.}
\label{histo_widths} 
\end{figure}

\begin{figure}
\centering
\includegraphics[width=0.45\textwidth]{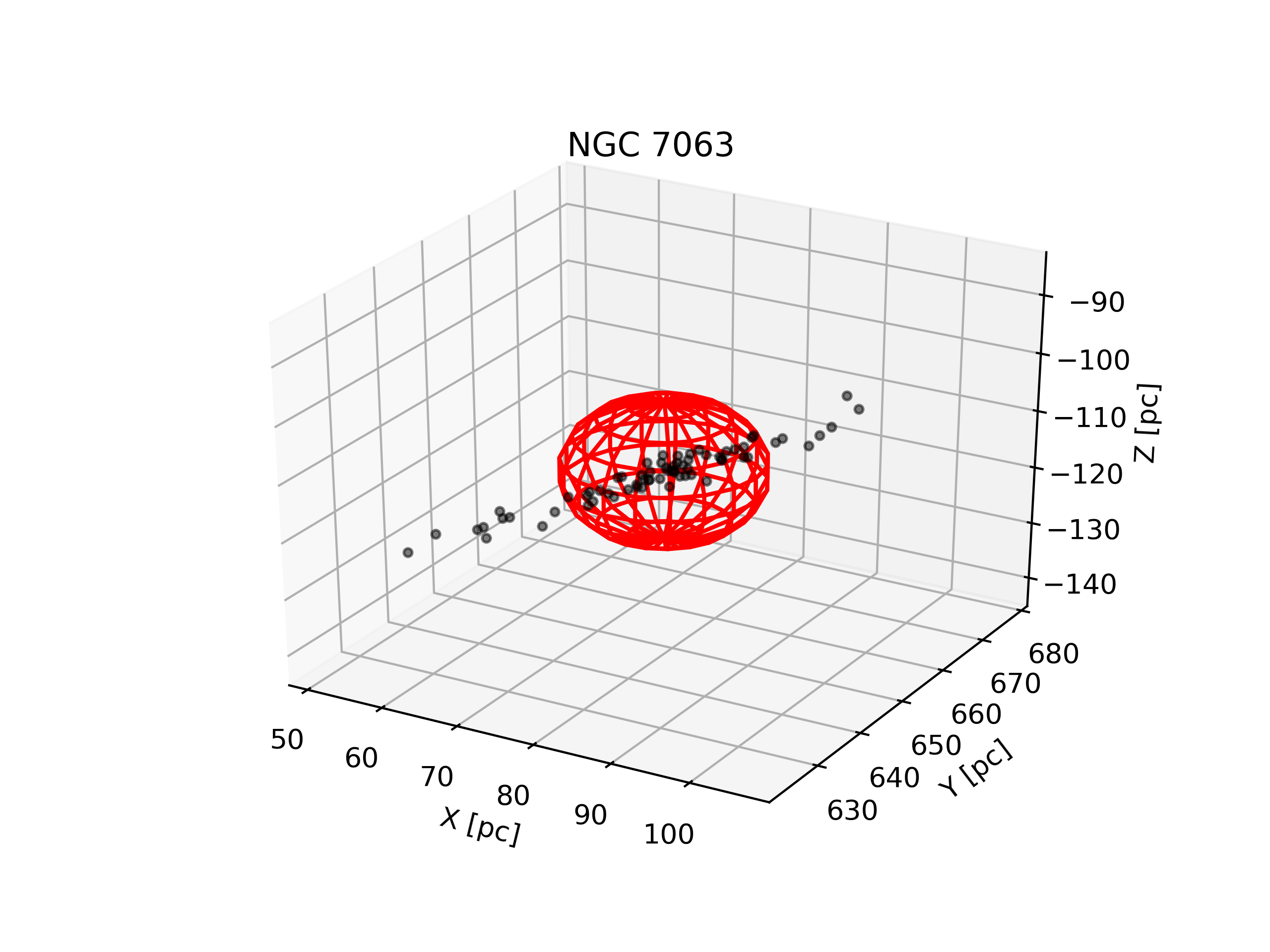}
\caption{The 3D structure of NGC 7063 at $d$\,=\,659(4)\,pc for all stars with a membership probability larger
than 50\%. The elongation is due to uncertainties in the observed parallaxes. 
Included is a sphere with a radius of 12\,pc centred at the
star cluster.}
\label{NGC_7063} 
\end{figure}  

\subsection{Characterizing clusters in three spatial dimensions} \label{3D_results}

Here, we follow the approach presented by \citet{2021arXiv210707230P}, which we briefly summarise.
The starting point are the rectangular Galactic coordinates [$X\,Y\,Z$] using the spherical 
Galactic coordinates $d$ (distance from the Sun), $l$ (Galactic longitude), and $b$ (Galactic latitude)

\begin{equation}
\begin{aligned}
X &= d \, \cos{l} \, \cos{b} \,\,,\\
Y &= d \, \sin{l} \, \cos{b} \,\,,\\
Z &= d \, \sin{b} \,\,.
\end{aligned}
\end{equation} 

To describe a cluster in a spatial 3D space, we can assume spherical symmetry and only need to 
choose two coordinates. The first coordinate is the projected distance, denoted as $d_1$. 
This can be calculated by projecting the actual distance $d$ of a star onto the vector 
that points along the line of sight to the cluster's centre.

\begin{equation}
d_1 = d \, \cos(\phi)
\end{equation}

where the variable $\phi$ represents the angle between the reference vector 
(which is located at the centre of the cluster) and the vector directed toward the 
specified star. It is important to note that the value of $d_1$ is nearly identical
to the value of $d$. Therefore, we will use $d$ as the measure of distance for this work. 
The second coordinate, $d_2$, represents the projected width and can be calculated similarly.

\begin{equation}
d_2 = d \, \sin(\phi).
\end{equation}

We analyzed the distribution of Galactic coordinates $l$ and  $b$, which are determined with high 
precision. We calculated the angular size of a cluster, $\phi_{\textrm{clust}}$, as 
the mean of the standard deviations of these two coordinates. The absolute size of the 
cluster can then be determined by its projected width, $d_2$. To find the typical 
spatial diameter of a cluster, we can examine the distribution of these projected widths, 
which can be done by calculating the median and the median standard deviation, for example.
As the next step, we can analyse the distribution of projected widths. Figure \ref{histo_widths}
shows the histogram of the projected widths, which peaks at about 4.1(0.9)\,pc, respectively.
Moreover, the upper limit of absolute cluster diameters of about 25\,pc \citep{2024A&A...686A..42H} is consistent with this. This upper limit arises from the size of the initial molecular cloud 
from which the clusters are formed, along with the dissipation due to the differential 
rotation of the Milky Way \citep{Joshi2016}.

It is important to note that the size of a cluster, as determined from projected widths, is 
heavily influenced by the method used to assess the cluster membership probabilities. 
Old star clusters have undergone significant dynamical evaporation. 
We conclude that the widths we calculated for the clusters may contain some systematic errors. 
These errors are likely negligible for younger clusters, but become significantly larger for 
older ones. Consequently, we anticipate that this could affect the distribution shown 
in Fig.~\ref{histo_widths}, resulting in a slight enhancement of the tail toward larger 
projected widths while simultaneously lowering the peak at lower values.

In our final step, we examined the three-dimensional characteristics of star clusters 
located within 2 kpc. To achieve this, we mapped the coordinates $d_1$ (or $d$) and $d_2$ 
of all individual members, creating a distribution that can be illustrated 
using histograms. Utilising data from $Gaia$ DR3 along with the distances $d$, we were 
able to generate scatter plots based on the [\(X, Y, Z\)] coordinates of the cluster members.

For more distant clusters, we would anticipate a needle-like 3D structure due to the significant 
errors in distance measurements. Interestingly, we can clearly observe this structure even for 
distances less than 750 pc. In contrast, nearby, densely populated clusters such as NGC 2632 
(with a radius of less than 250 pc) exhibit only a slight elongation along the line of sight. 
This observation highlights the limitations of even the best available data in studying the 
3D structures of open clusters.

In Fig.~\ref{3D_four_clusters}, we present the spatial distributions for four clusters: 
NGC 1039 (distance of 509(4) pc), NGC 1528 (1038(6) pc), NGC 2632 (192(1) pc), and 
NGC 5823 (1756(14) pc). The anticipated needle-like structure along the line of sight 
is clearly visible. It is noteworthy that the apparent members of NGC 5823 are 
distributed approximately 165 pc around the Galactic disk.

We aim to demonstrate how the results impact the estimates of the 3D radii of individual clusters. 
We fitted each cluster with a sphere that encompasses 50\% of its population, using the median value 
of the [X, Y, Z] coordinates as the cluster centre. These spheres can be compared to a reference 
sphere with a radius of 12 pc, allowing us to identify the sphere that is either smaller than or 
approximately equal to 12 pc and is the most distant from the Sun. 

In Fig. \ref{NGC_7063}, we present the case of NGC 7063, which is located at a distance of 
659(4) pc from the Sun. All stars with a membership probability greater than 50\% are included 
in this analysis. It is notable that the distribution of these members is not spherical but 
rather needle-like.

This analysis serves as an upper limit for the distance within which we can reasonably fit 
spheres to open clusters. However, it is important to emphasise that the true internal 3D 
structure of an open cluster cannot be accurately assessed. This limitation primarily arises 
from the elongation caused by observational uncertainties. Additionally, there is a secondary 
factor to consider: the bias introduced by selecting member stars based on line-of-sight observations. 
In the future, algorithms designed to identify members of more distant star clusters should 
convert astrometric data into the [X, Y, Z] space, rather than relying on direct line-of-sight 
distances. This approach would enable the search for members in a three-dimensional space
around the cluster centre, thus minimising potential (albeit small) selection effects.

One well-known phenomenon is the tidal tails of star clusters. Star clusters are expected to 
have a finite lifetime due to dynamical evaporation \citep{2024ApJ...970...94K}. 
This is especially true for older open clusters, where determining the total mass at birth is 
challenging due to the loss of members over hundreds of millions of years. Accurately 
estimating this parameter is crucial for constraining the star formation rate in the Milky Way. 

Identifying these tidal tails can be complicated, and one must calculate the space velocities of 
stars toward a convergent point. However, since radial velocity measurements are lacking for most 
stars, studies rely solely on tangential velocity criteria. It's important to note that the needle-like 
structures do not represent tidal tails; rather, they stem from uncertainties in the observed 
parallaxes of the individual members of star clusters.

\section{Comparison with the literature} \label{comparison_literature}

The last three years saw a flood of papers presenting total masses, core radii, and other cluster
parameters
\citet{2023A&A...672A.187J,2023MNRAS.525.2315A,2024A&A...685A..33R,2024A&A...686A..42H,2025AJ....169..115W}.
Different papers used different names for the same parameters, see the Jacobi radius in 
\citep{2024A&A...686A..42H}. Unfortunately, the different papers presented widely different
values for the same cluster. Just one example, the total mass published in the above-mentioned 
papers for NGc 2682 range between 1327 and 7437 M$_\odot$, or a factor of six. The same is true for
the very different definitions of clusters' radii. 
It is out of the scope of this paper to find reasons for these offsets. 
So, our values always fall somewhere in between any published values. We have checked for possible
correlations of the offsets with the age and metallicity, but found none.

The reader must be aware that
the current status of these specific cluster parameters is very unsatisfactory and needs special
attention in the future. 
For sure, we are now in the lucky position of having very good photometric and astrometric 
data based on the $Gaia$ satellite mission. It is now time to agree how to get reliable 
cluster parameters and how to test them with, for example, benchmark open
clusters.

\section{Conclusions} \label{conclusions}

The stars in the 7167 open clusters were processed in Python with custom code using the existing Scipy package. The density profiles of the clusters showed the diversity of their shapes and star distribution. Approximately 53\% of the clusters have GOF values above 0.9, indicating good agreement between the observed density profile and the adopted exponential model. In contrast, clusters with low or zero GOF values (approximately 23 \% of the total) show either a small number of stars, an irregular distribution, or a mismatch of the data with the density profile model used. These results are consistent with the influence of dynamical processes, such as gravitational interactions and tidal forces, on the shapes and distributions of stars within individual clusters. The elongated and irregular shapes observed, for example, in the Beta Tucanae Group or ASCC 85, may be the result of these processes, while clusters with high GOF values, such as NGC 362, are characterised by a strong central concentration of stars and a stable structure.

The analysis of open clusters with available stellar masses provides insight into their mass profiles. Mass profiles were calculated at mass percentiles (2.5, 16, 50, 84, and 97.5), allowing us to assess the distribution of stellar masses in each cluster. The results showed that the total mass of the clusters increases with higher percentiles as they include more massive stars, while lower percentiles correspond to lower-mass stars, as we would expect.
The cluster with the largest total mass (Theia 3623, 9019 M$_\odot$) includes 3086 members and shows an exponential decrease in mass with distance, indicating effective mass segregation. In contrast, the least massive cluster (HSC 1941, 6.2 M$_\odot$) with only 10 members lacks a distinct mass profile, probably due to the low number of stars.
An interesting example is the cluster NGC 6231, which contains 646 members and has a total mass of 7041 M$_\odot$. This indicates the presence of mostly very massive stars. Given its young age (2  -7 million years), many of its massive stars are expected to remain on the main sequence. Furthermore, the mass profile shows sharp fluctuations, indicating the presence of less massive stars. The density profile exhibits an exponential decrease with distance from the cluster centre, while the spatial distribution of the stars appears approximately spherical.
The older cluster NGC 6791 (4 Gyr) contains 4193 members with a total mass of 4870 M$_\odot$. Its exponential decrease in mass with distance and its spherical density profile are consistent with a higher degree of mass segregation, consistent with the cluster's age and the long-term gravitational interactions among its members.
Younger clusters, such as Alessi 43 (30 Myr), exhibit an irregular mass profile, suggesting a lack of mass segregation. In addition, this is the cluster containing the most massive star in our data sample (63.1 M$_\odot$). In contrast, the Alessi 13 cluster, with 167 members and a total mass of 101.9 M$_\odot$, contains mostly light stars, with the least massive stars reaching only 0.09 M$_\odot$.
Of particular note are the clusters CWNU 1044, HSC 749 and HSC 939, whose mass and density profiles show different structural properties. CWNU 1044 shows an irregular decrease and increase in mass with distance from the cluster centre, probably due to its elongated shape. HSC 749 contains two distinct, separated regions with stars, and HSC 939 has stars distributed in three clusters around the cluster centre.
Overall, the dynamical and structural properties of clusters, such as mass segregation, density profile, and mass distribution, appear to be related to their age, number of members, and total mass. Middle-aged clusters show more pronounced segregation and a more regular distribution of stars, while younger and less massive clusters often lack systematicity in their mass and density profiles.

We investigated the limitations of $Gaia$ DR3 for studying open clusters. Due to uncertainties in observed parallaxes, many clusters exhibit elongated, needle-like shapes rather than a spherical form. With the currently available data, we can effectively study the diameters of open clusters up to about 2 kpc using a statistical approach. Our findings align with existing models, showing that all clusters have diameters of less than 25 pc, with a peak diameter observed between 3 and 5\,pc. However, it is important to note that these results depend heavily on the method used to determine the cluster membership probabilities. Additionally, we found that individual open clusters located beyond 500\,pc should not be included in 3D studies when using the most commonly used parallax-to-distance transformation methods.

Still, no common scale of cluster characteristics is available in the literature. The published
values for the total masses and radii significantly deviate from each other without any obvious
correlations. 

Our results further demonstrate the need for a homogeneous scale of open-cluster parameters. Published estimates of cluster masses and radii often differ substantially, highlighting the importance of consistent analysis methods.

\begin{acknowledgements}

This work was supported by the grant GA{\v C}R 23-07605S.
This research has made use of the SIMBAD database, operated at CDS, Strasbourg, France and of the WEBDA database, operated at the Department of Theoretical
Physics and Astrophysics of the Masaryk University.
This work has made use of data from the European Space Agency (ESA) mission 
{\it Gaia} (\url{https://www.cosmos.esa.int/gaia}), processed by the {\it Gaia} Data 
Processing and Analysis Consortium (DPAC, \url{https://www.cosmos.esa.int/web/gaia/dpac/consortium}). 
Funding for the DPAC has been provided by national institutions, in particular the institutions
participating in the {\it Gaia} Multilateral Agreement. 

\end{acknowledgements}

\bibliographystyle{aa} 
\bibliography{paper.bib}




\end{document}